\documentclass[11pt]{article}
\usepackage{geometry}
\usepackage{amsmath,mathtools}
\usepackage{amsfonts}
\usepackage{bm}
\usepackage{amsthm}
\usepackage{enumerate}
\usepackage{esint}
\usepackage{color}
\usepackage [english]{babel}
\usepackage [autostyle, english = american]{csquotes}
\usepackage[T1]{fontenc}
\usepackage[utf8]{inputenc}
\usepackage{authblk}
\usepackage{graphicx}
\usepackage{subcaption}
\usepackage{titlesec}
\usepackage{tocloft}

\titleformat*{\section}{\centering \large\bfseries}
\titleformat*{\subsection}{\centering \large\itshape}

\MakeOuterQuote{"}

\allowdisplaybreaks

\newcommand{\beq}{\begin{equation}}
\newcommand{\eeq}{\end{equation}}
\newcommand{\beqs}{\begin{equation*}}
\newcommand{\eeqs}{\end{equation*}}
\newcommand{\bal}{\begin{align}}
\newcommand{\eal}{\end{align}}
\newcommand{\bals}{\begin{align*}}
\newcommand{\eals}{\end{align*}}

\theoremstyle{definition}

\title{Patterning nonisometric origami in nematic elastomer sheets}
\author[a]{Paul Plucinsky} 
\author[b]{Benjamin A. Kowalski} 
\author[b]{Timothy J. White}
\author[c]{Kaushik Bhattacharya}
\affil[a]{Aerospace and Engineering Mechanics, University of Minnesota}
\affil[b]{Materials and Manufacturing Directorate, Air Force Research Lab}
\affil[c]{Engineering and Applied Sciences, California Institute of Technology}
\begin{document}
\maketitle

\begin{abstract}
Nematic elastomers dramatically change their shape in response to diverse stimuli including light and heat.  In this paper, we provide a systematic framework for the design of complex three dimensional shapes through the actuation of heterogeneously patterned nematic elastomer sheets.  These sheets are composed of \textit{nonisometric origami} building blocks which, when appropriately linked together, can actuate into a diverse array of three dimensional faceted shapes.  We demonstrate both theoretically and experimentally that the nonisometric origami building blocks actuate in the predicted manner, and that the integration of multiple building blocks leads to complex, yet predictable and robust, shapes.   We then show that this experimentally realized functionality enables a rich design landscape for actuation using nematic elastomers.  We highlight this landscape through examples, which utilize large arrays of these building blocks to realize a desired three dimensional origami shape. In combination, these results amount to an engineering design principle, which provides a template for the programming of arbitrarily complex three dimensional shapes on demand.
\end{abstract}

\section*{Introduction}

The seamless integration of function and form promises to spur innovation in technologies ranging from MEMS and NEMS devices (e.g., with novel electrical, electromagnetic and energy functionality), reconfigurable and soft robotics, wearable electronics, and compliant bio-medical devices \cite{retal_mrs_16,xetal_science_15,hetal_pnas_10,fetal_science_14,cm_sr_14,kr_advMat_08}.  This integration can be facilitated by incorporating soft active materials into thin or slender structures to program complex three dimensional shapes not easily achieved by conventional means of manufacturing.  This is not without its challenges: The coupling of nonlinearities\textemdash at the material level and at the structural level\textemdash makes a salient and general theory of design with these systems a trying task.  Even more, bridging the gap between an idealized theory and what is possible (and practical) experimentally offers a different, but equally important, set of challenges.  In this work, we address these challenges in the context of active patterned sheets capable of localized anisotropic distortion under stimuli.  Specifically, we identify the framework for designing these sheets to achieve desired faceted shapes through actuation, and we leverage recent advances in the localized mechanical response of nematic elastomers to realize the designed shape control.

\begin{figure}[!t]
\includegraphics[width = 6.5in]{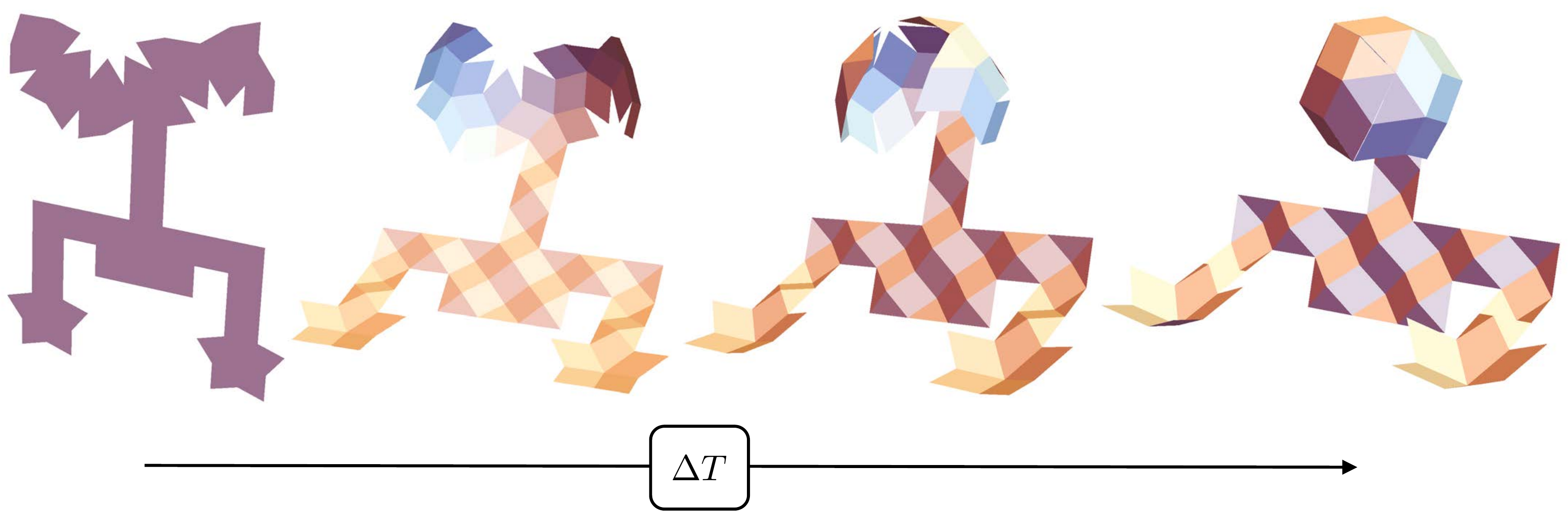}
\caption{A "humanoid" soft robot.}
\label{fig:SoftRobot}
\end{figure}

Nematic elastomers couple the elasticity of a soft and highly deformable polymer network with the orientational ordering of rod-like liquid crystal molecules (mesogens), which are either incorporated into the main chains of the network or pendent from them.  This coupling results in a solid with dramatic shape-changing response to temperature change and other stimuli \cite{wt_book_03,wb_nm_15}: At low temperatures, the liquid crystals prefer being aligned (in some average sense), with the orientation of this alignment described locally by a unit vector $\mathbf{n}_0$ called the director.  Upon heating, however, thermal fluctuations suppress this tendency towards order, resulting in a nematic to isotropic phase transformation within the solid.   This gives rise to a strongly anisotropic macroscopic deformation since the polymer network is intrinsically coupled to this order.  Typically, the elastomer contracts along the director and expands transversely.

Building on key ideas in the study of non-Euclidean plates \cite{kes_science_07,esk_jmps_09,se_sm_10}, Modes et al.\ \cite{mbw_pre_10,mbw_prsa_11} recognized that by programming the director \textit{heterogeneously} in the plane of a thin sheet, stimulation results in inhomogeneous shape-change that in turn drives complex three dimensional shapes.  Indeed, they showed theoretically how patterning azimuthal and radial director profiles about a defect enabled conical and saddle-like shapes upon actuation.  These were later realized experimentally\textemdash first by de Haan et al.\ \cite{detal_acie_12} in nematic glass sheets and later by Ware et al.\ \cite{wetal_science_15} in nematic elastomers.   Even more, the latter work made possible the prescription of an arbitrary planar director profile in a thin sheet (i.e., through \textit{voxelated} LCEs), bringing questions of designable actuation to the forefront.    Further, since the entire sheet participates in the actuation, it is extremely robust and capable large actuation force and energy.  This has motivated the study of other more complex patterns  \cite{mw_pre_11,ask_prl_14,m_pre_15,mwww_prsa_16,plb_pre_16,plb_arma_17}. 

Of particular interest to this work is the class of {\it nonisometric origami} patterns, in which the director is piecewise uniform in the plane.  These enable the design of complex faceted (origami) shapes from simple building blocks  \cite{plb_pre_16,plb_arma_17}. However, there are two significant open issues which need to be addressed before these ideas can be used to reliably to realize designed shape control.  

\begin{figure*}[!t]
\centering
\includegraphics[width = 6.5in]{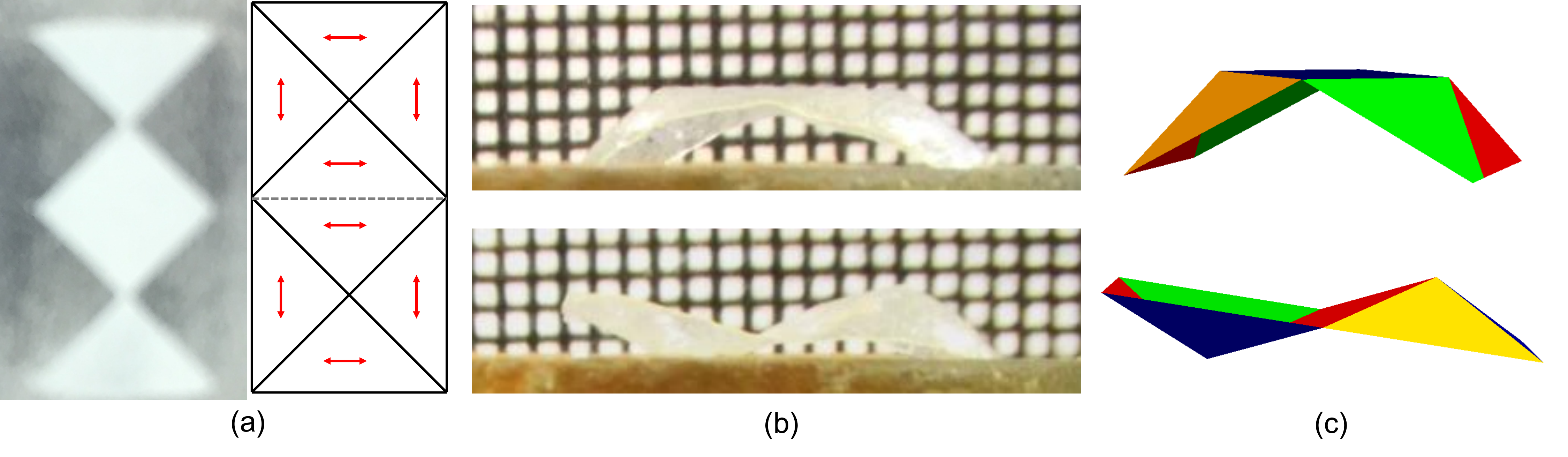}
\caption{(a) The design and experimental realization for two linked symmetric four-faced building blocks (right and left respectively).  (b) The sheet actuates into either of two degenerate and mechanically bi-stable configurations.  (c)  These configurations are predicted by theory, as discussed below. }
\label{fig:TwoPyramid}
\end{figure*}

First, all the patterns described in the literature \cite{mbw_pre_10,mbw_prsa_11,detal_acie_12,wetal_science_15,mw_pre_11,ask_prl_14,m_pre_15,mwww_prsa_16,plb_pre_16,plb_arma_17} have multiple, energetically degenerate actuated configurations.  For example, the cone can actuate either up or down. This is compounded by the fact that, as patterns are made more complex, the number of multi-stable configurations increases.  We illustrate this in Figure \ref{fig:TwoPyramid}, showing a nonisometric origami pattern comprising a joined pair of actuating pyramids.  (The pyramid design is discussed in more detail later on).  Briefly, this sample exhibits two non-trivially distinct shapes, which are both stable as the pyramids are free to actuate either up or down: If both actuate up (or down), then the sample realizes the shape depicted at the top in \ref{fig:TwoPyramid}(c).  Alternatively, if one actuates up and the other down, then the sample realizes the bottom shape.  There is nothing inherent to this design to break the up-down symmetry, and so both configuration are observed in the single sample shown.  Thus, in order to deterministically actuate a patterned sheet into a \textit{single} desired configuration, this degeneracy must be broken, ideally in a way that allows us to arbitrarily program individual folds to actuate up or down as prescribed.

We aim to introduced directional bias in the natural deformations of these sheets to break the degeneracy.  One approach is  to introduce a hierarchical rotation in the director profile\textemdash where the planar director varies through the thickness of the sheet\textemdash in what is called a \textit{twisted nematic profile}.  These profiles are known to generate a spontaneous curvature upon actuation which results in complex bending deformations both in isolated strips \cite{metal_afm_05,wmc_prsa_10,setal_pnas_11} and when integrated into larger patterned sheets \cite{metal_pre_12,fetal_rsc_15,getal_sr_17}.   However, twisted nematics induce a spontaneous Gauss curvature, which can lead to undesirable anti-clastic bending.  Even further, Gauss' Theorema Egregium states that a change in Gauss curvature is necessarily accompanied by a stretch.  Consequently, this mode of actuation always results in mechanical frustrations in the sheet that influence the robustness of actuation \cite{wetal_prsa_10}.  So, this strategy has proven effective only when applied in narrow regions.  Unfortunately, this restricts the actuation force and energy.

Our key idea is a hybrid approach in which we modify nonisometric origami so that large regions of uniform director are joined by narrow boundary regions of slight nematic twist.  However, this is subtle and we provide explicit and quantitative design guidance for introducing the bias but also suppressing unwanted anti-clastic bending at the boundary regions.  We then incorporate this bias into simple building blocks of nonisometric origami, and show that this is an experimentally accessible design strategy; specifically, in monolithic sheets of uniform composition, varying only in director orientation in the faces and hinges.  With these, we achieve a robust actuation, with large actuation force and energy.

The second largely open issue concerns broad strategies for the design complex shapes:  While such strategies are emerging in the recent literature for radial and smooth surfaces \cite{wm_arXiv_17, aetal_arXiv_17}, these require very precise control over the director field in the patterned sheet.  This can be difficult to achieve experimentally.  Our key idea, however, exploits the capability of nonisometric origami to create \textit{simple building blocks} that can be composed systematically to achieve faceted shapes of arbitrary complexity using monolithic but heterogeneously patterned sheets. This provides a design landscape with vast potential; one such example is highlighted in Figure \ref{fig:SoftRobot}.  

\subsection*{Materials and methods}

Patterned LCE films are fabricated following procedures described in \cite{wetal_science_15,ketal_sm_17}.  In short, glass slides are spin-coated with a commercial photoalignment material (PAAD-22, BEAM Co).  Photoalignment is performed by sequential exposures to a focused 445 nm laser spot.  By adjusting the laser's linear polarization at each successive exposure location, we build up the desired complex alignment pattern.  This pattern is then permanently fixed by spin-coating with a thin layer of mesogenic monomer (RM257, Synthon, 3 wt solution in dichloromethane) and flood-curing.  Finally, two slides are glued together to form an alignment cell, using glass beads to ensure a 50 micron cell spacing.

A previously reported one-pot LCE formulation \cite{wetal_acsml_15} is then capillary filled into the alignment cell in the isotropic phase.  Upon cooling to the nematic phase, it is flood-cured to form a free-standing monolithic elastomer film.  Heat-stimulated shape change is quantitatively measured using a structured-illumination optical 3D scanner with micron-scale height resolution (Keyence VR-3200).

\subsection*{Notation}  We will be dealing with  vector quantities, both in two and  three dimensions.  Sometimes these will be in the same equation. Thus, we use tilde as a means of distinguishing these quantities.  For instance, we use $\mathbf{v}$ for a vector on $\mathbb{R}^3$ and $\tilde{\mathbf{v}}$ for a vector on $\mathbb{R}^2$; we define the standard basis on $\mathbb{R}^3$ as $\{\mathbf{e}_1, \mathbf{e}_2, \mathbf{e}_3\} \in \mathbb{S}^2$ and the standard basis on $\mathbb{R}^2$ as $\{ \tilde{\mathbf{e}}_1, \tilde{\mathbf{e}}_2\} \in \mathbb{S}^1$; we set the Cartesian coordinates $\mathbf{x} := x_1 \mathbf{e}_1 + x_2 \mathbf{e}_2 + x_3 \mathbf{e}_3$ and $\tilde{\mathbf{x}} := x_1 \tilde{\mathbf{e}}_1 + x_2 \tilde{\mathbf{e}}_2$; we define the three dimensional gradient as $\nabla$ (with respect to $\mathbf{x}$) and the planar  gradient as $\tilde{\nabla}$ (with respect to $\tilde{\mathbf{x}}$), etc.   Here and throughout, the set $\mathbb{R}^n$ denotes the $n$-dimensional real space with $\mathbb{S}^{n-1}$ denoting the set of unit vectors on $\mathbb{R}^n$.

\section*{Results}

\subsection*{Metric constraint}

The fundamental idea of shape morphing is to write a heterogeneous pattern of director field on a thin sheet that when actuated creates a pattern of spontaneous stretch that is not compatible in the plane, thereby leading the sheet to a complex three-dimensional shape.  This can be described using a metric constraint \cite{esk_sm_13,ask_prl_14, plb_pre_16} that relates the three-dimensional deformation $\mathbf y$ of the mid-plane of a  thin sheet $\omega \subset \mathbb{R}^2$ to the induced metric due to the actuation (described by a parameter $r\equiv r(\Delta T) >0$ with $r = 1$ prior to actuation and $\in (0,1)$ for heating) and a director pattern $\mathbf{n}_0 \colon \omega \rightarrow \mathbb{S}^2$:
\begin{align}\label{eq:metric2D}
(\tilde{\nabla} \mathbf{y})^T \tilde{\nabla} \mathbf{y}  = r^{-1/3} ( \mathbf{I}_{2 \times 2} + (r-1) \tilde{\mathbf{n}}_{0} \otimes \tilde{\mathbf{n}}_0) = : \tilde{\bm{\ell}}_{\mathbf{n}_0},
\end{align}  
where $\tilde \nabla$ is the in-plane gradient and $\tilde{\mathbf{n}}_0 := ( \mathbf{n}_0 \cdot \mathbf{e}_1, \mathbf{n}_0 \cdot \mathbf{e}_2) \in \mathbb{R}^2$ is the projection of the the director onto the plane.  Note, this formulation is for soft and lightly cross-linked nematic elastomers (which are nearly incompressible), but it is easily adapted to nematic glasses \cite{wmc_prsa_10,ask_prl_14,m_pre_15} for which the incompressibility assumption breaks down.  Further, one can justify this metric constraint rigorously \cite{plb_arma_17} on the basis of energy minimization of the well-established free energy of nematic elastomers \cite{btw_pre_93,wt_book_03}.  Finally, much of the work thus far has focused on identifying specific patterns ${\mathbf n}_0$ that are consistent with interesting deformations (shapes) $\mathbf y$.

There are two broadly open issues:  The first is degeneracy, as there often are many deformations that satisfy the metric constraint \eqref{eq:metric2D} for a given director pattern, and the second is the systematic design of complex shapes.  In this paper, we propose a strategy that involves {\it nonisometric origami} where both the director and the deformation gradient are piecewise uniform to address both these issues.

\subsection*{Nonisometric origami}

\begin{figure}
\centering
\includegraphics[width = 5in]{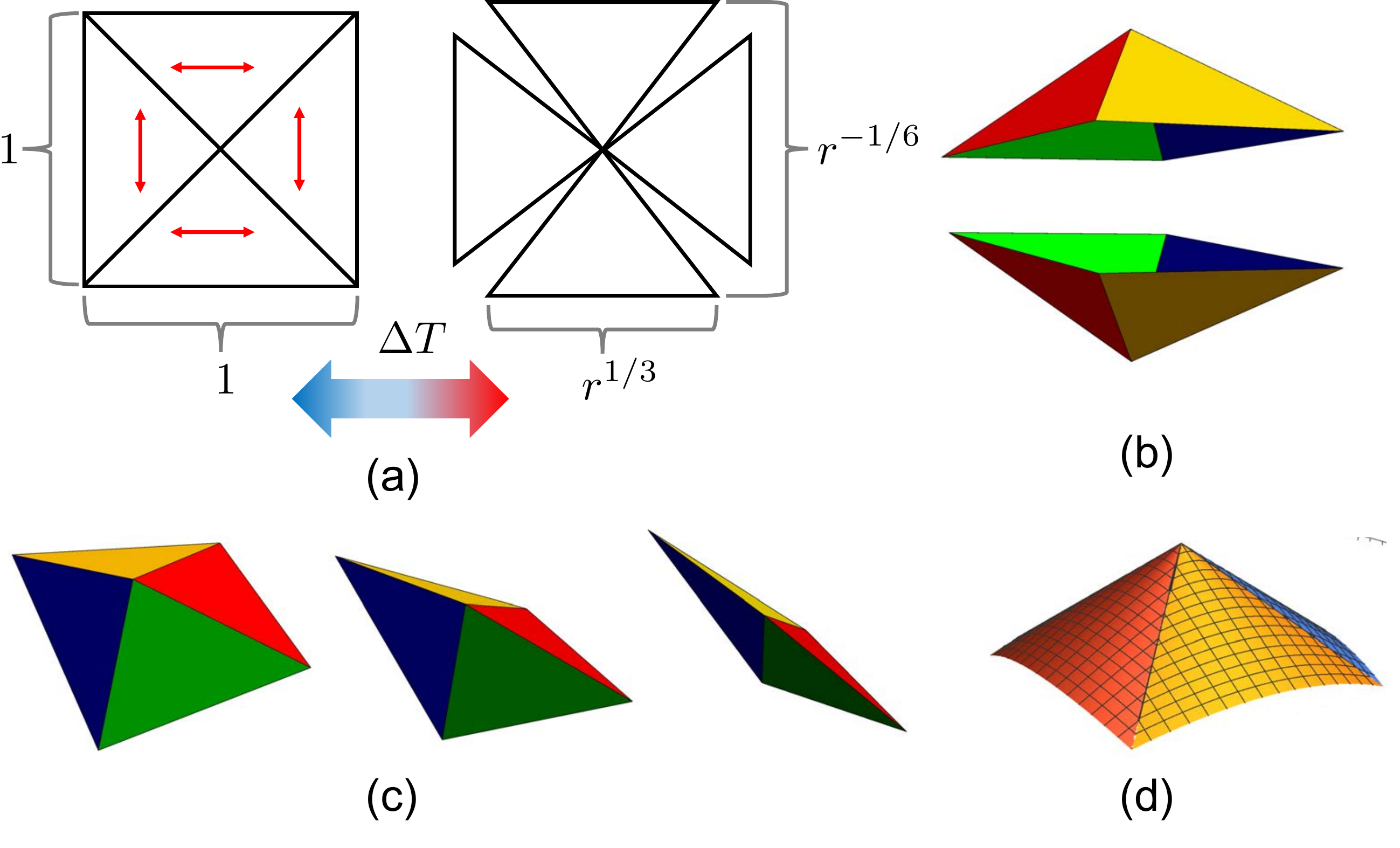}\\
\caption{A simple example of a nonisometric origami.  (a) The shape change induced by actuation of this patterned design cannot be accommodated by a planar deformation of the sheet.  Alternatively, the sheet can form a pyramid to realize this shape change (metric).  However, there are many degeneracies: (b) the pyramid can actuation up or down; (c) the folding angles need not be equal; (d) each triangle can curve so that the actuation is conical.}
\label{fig:Degen}
\end{figure}

We introduce nonisometric origami  through a simple example where a flat sheet is actuated into a pyramid.  Consider a unit square sheet made of equal triangles of base $1$ and height $1/2$ as depicted Figure \ref{fig:Degen}(a).  In each triangle the director is programmed in the plane and is parallel to the base of the triangle.  Thus upon heating, each triangle desires a spontaneous contraction $r^{1/3}$ along its base and an expansion $r^{-1/6}$ along its height, consistent with \eqref{eq:metric2D}.  Notice, though, that it is not possible to satisfy this shape changing distortion in the plane without breaking up sheet along one of the interfaces distinguishing the four triangles.  Alternatively, notice that by rotating each of these triangles out-of-plane, we can form a pyramid to accommodate the heterogeneous shape changing distortion while keeping the interfaces intact.  That is, through the pyramid actuation, we obtain a deformation which satisfies the induced metric in \eqref{eq:metric2D}.   Thus, we realize a three-dimensional faceted shape by patterning the directors to be piecewise constant in polygonal regions.  This is an example of \textit{nonisometric origami}.   But this is only a simple example and one can compose this and other simple examples\footnote{See supplementary material.} as building blocks to achieve more complex shape \cite{plb_pre_16,plb_arma_17,p_thesis_2017}.  Thus, nonisometric origami is a promising route to potentially arbitrary complexity in shape.  

However, the examples suffer from degeneracy, which limits their utility as building blocks.   We highlight this degeneracy with Figure \ref{fig:Degen}(b-d) for the simple design to obtain the pyramid.  Notice, the pyramid can go up or down.  Further, the folding angles do not all have to be equal, as the symmetric pyramid can be deformed without stretching the triangles into a range of pyramidal shapes.  Finally, the faces do not have to be flat; instead each triangle can be deformed without stretch into a conical strip.

\begin{figure*}[t]
\centering
\includegraphics[width = 6.5in]{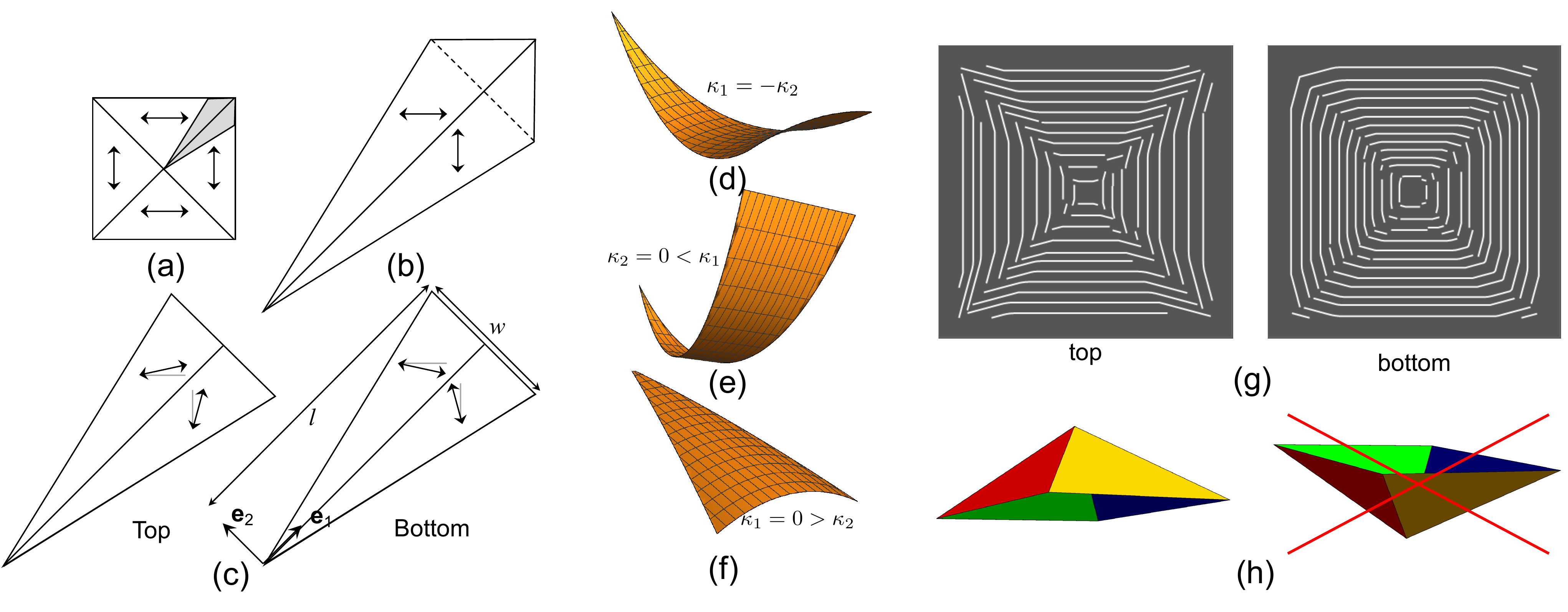}
\caption{The design of a pyramid with bias.  (a) The director pattern of non-isotropic origami pyramid and wedge-shaped region around an edge.  (b) Details of the wedge-shaped domain highlighted in (a).  (c) The modification of the origami director pattern to introduce a twist.  (d-f) The possible actuated shape on heating of the triangular specimen shown in (c).  (g) The director pattern for a biased pyramid. (h) The up pyramid is preferred over the down pyramid. }
\label{fig:DesignRidge}
\end{figure*} 

\subsection*{Strategy to bias the actuation}  

We seek a strategy to select amongst the various degenerate shapes, and the basic idea is relatively simple.  These shapes all involve the same stretch, but differ by the curvature.\footnote{Note that the folding angle in idealized origami actuation will have finite curvature in reality.}  Therefore, we can select a shape by a preferred curvature at each point.  We do this by introducing slight twist in the nematic director in a monolithic sheet in the regions which form the hinges of the origami\footnote{This is achieved experimentally by patterning the two sides of the alignment cell slightly differently (see the material and methods section).}.  These twist profiles introduce a preferred spontaneous curvature upon actuation, but this requires careful study.  Indeed, if the twist profile is not properly designed, then actuation can lead to anti-clastic bending (i.e., negative Gauss curvature) and, as a consequence, undesirable stretch which would compete with the designed nonisometric origami.   

To understand the various issues involved, we return to our example of a pyramid and study in some detail the narrow wedge around an edge.  This wedge is highlighted in Figure \ref{fig:DesignRidge}(a) and isolated in Figure \ref{fig:DesignRidge}(b).  The director profile according to nonisometric origami is also marked in the figure.  It is convenient to analyze the triangular or wedge-like region indicated by the dashed line.  We introduce a twist in this region as shown in Figure \ref{fig:DesignRidge}(c).  On the top surface of the sheet, we rotate the director slightly towards being parallel to the edge, while on the bottom surface we rotate the director slightly in the opposite sense towards being perpendicular to the edge.  In other words, we introduce a small twist while retaining the original nonisometric origami design in the mean profile averaged through the thickness.  In what follows, we show that the wedge  deforms in one of three possible ways shown in Figure \ref{fig:DesignRidge}(d-f) depending on a non-dimensional parameter $g_f$ defined in (\ref{eq:geomFact}).  If this parameter is small, then the wedge deforms into the anti-clastic shape with two opposite curvature shown Figure \ref{fig:DesignRidge}(d);  conversely if the parameter is large, then one of the curvatures is suppressed and it deforms in one of the two ways shown in Figure \ref{fig:DesignRidge}(e,f).  We incorporate this understanding of the deformation of the wedge to the actuation of the pyramid in the next section.

To establish the result, let $\langle \mathbf{n}_0\rangle$ denote the director distribution according to the nonisometric origami design shown in Figure \ref{fig:DesignRidge}(b) and $\langle \mathbf{n}_0\rangle^\perp$  direction perpendicular to it in the plane.  Then, the twist profile of Figure \ref{fig:DesignRidge}(c)  can be described as\begin{align}
\mathbf{n}_0^h(\mathbf{x}) =  \cos\big( \frac{\tau x_3}{h} \big) \langle \mathbf{n}_0\rangle (x_2)   + \sin \big(\frac{\tau x_3}{h} \big)  \langle \mathbf{n}_0\rangle^{\perp} (x_2).
\end{align}
in the coordinate system shown in Figure \ref{fig:DesignRidge}(b).   This director profile leads to a spontaneous stretch described by the square-root of the fully three-dimensional metric  
\begin{equation} 
\begin{aligned}\label{eq:localStretch}
\bm{\ell}_{\mathbf{n}_0^h}^{1/2}(\mathbf{x}) &:= r^{-1/6}(\mathbf{I}_{3\times3} + (r^{1/2}-1) \mathbf{n}_0^h(\mathbf{x}) \otimes \mathbf{n}_0^h(\mathbf{x})) \\
& \ = r^{-1/6}\Big( \mathbf{I}_{3\times3} - \epsilon \big( \mathbf{A}_{S}(x_2) + \big(\tau x_3/h \big) \mathbf{A}_B  +  O(\tau^2) \big)\Big),
\end{aligned}
\end{equation}
where the second equality is obtained by a Taylor expansion in the twist angle $\tau$.  Here, 
\begin{equation}
\begin{aligned}\label{eq:ASAB}
&\mathbf{A}_S(x_2) =  \langle \mathbf{n}_0 \rangle (x_2) \otimes \langle \mathbf{n}_0 \rangle(x_2),   \\
&\mathbf{A}_{B} = \mathbf{e}_1 \otimes \mathbf{e}_1  - \mathbf{e}_2 \otimes \mathbf{e}_2 \\
\end{aligned}
\end{equation}
are the spontaneous in-plane stretch and spontaneous curvature respectively and the parameter $\epsilon \equiv \epsilon(\Delta T)  := 1- r^{1/2}$ \footnote{Prior to actuation, $\epsilon = 0$ and the sheet is flat and undeformed.  Upon heating though, $\epsilon > 0$ and this is the driver for shape change within the sample.}.  Notice that the spontaneous curvature has eigenvalues (principal curvatures) of opposite sign, or  negative Gauss-curvature.  Due to Gauss' Theorema Egregium, it is not possible for the wedge to have both uniform stretch and uniform negative Gauss-curvature.  Thus, this twist nematic pattern will be internally stressed, and further analysis is required to understand the nature of equilibrium.  The basic idea in what follows is that the anti-clastic bending can be suppressed completely in favor of singly-bent ridges if the wedge is sufficiently wide.

We start from the total energy (according to the well-established free energy of nematic elastomers \cite{btw_pre_93,wt_book_03}), and study the asymptotics of this energy in thickness under the assumption that we have moderate stretch or $\epsilon \sim h^2$  similar to \cite{lmp_prsa_11}).  We obtain a von K\'{a}rm\'{a}n theory described by the energy
\begin{equation}
\begin{aligned}\label{eq:vkTheory}
&\mathcal{E}_{vK}( \tilde{\mathbf{u}}, v)=  \int_0^l \int_{-\tfrac{w x_1}{2l}}^{\tfrac{w x_1}{2l}} \Big( Q_2\big(\text{sym} \tilde{\nabla} \tilde{\mathbf{u}} + \frac{1}{2} \tilde{\nabla} v \otimes \tilde{\nabla} v + \epsilon \tilde{\mathbf{A}}_S(x_2)\big) \\
& \quad \quad \quad \quad \qquad  \qquad +  \frac{h^2}{12} Q_2\big( \tilde{\nabla} \tilde{\nabla} v - \frac{\epsilon \tau}{h}  \tilde{\mathbf{A}}_B\big)\Big) dx_1 dx_2,
\end{aligned}
\end{equation}
where $\tilde{\mathbf{u}}$, $v$  are the in-plane and out-of-plane displacements of the wedge, respectively, $Q_2(\tilde{\mathbf{F}}) = \mu (|\tilde{\mathbf{F}}|^2 + (\text{Tr } \tilde{\mathbf{F}})^2)$ with $\mu >0$ the shear modulus of the elastomer, and $\tilde{\mathbf{A}}_S(x_2)$ and $\tilde{\mathbf{A}}_B$ are the $2\times2$ principal minors of the tensors in \eqref{eq:ASAB}.  

To minimize this energy, we set $\tilde{\mathbf{u}}(\tilde{\mathbf{x}}) =\tilde{\mathbf{u}}^{\ast}(\tilde{\mathbf{x}}) + \epsilon( \tilde{\mathbf{x}} + |x_2| \tilde{\mathbf{e}}_1)$.  Notice that the latter term corresponds jump in the strain at the interface and satisfies the (linearized) metric constraint  $\text{sym} \tilde{\nabla} (\epsilon( \tilde{\mathbf{x}} + |x_2| \tilde{\mathbf{e}}_1)) = -\epsilon \tilde{\mathbf{A}}_S(x_2)$.   We then minimize the energy amongst the class of smooth polynomial displacements\footnote{The terms absent to leading order in this expansion vanish identically when accounted for in the energy minimization.}
\begin{equation}
\begin{aligned}
v(\tilde{\mathbf{x}}) &= \frac{\epsilon \tau}{h} \Big(\hat{\kappa}_1 x_1^2 + \hat{\kappa}_2 x_2^2 \Big), \\
\tilde{\mathbf{u}}^{\ast}(\tilde{\mathbf{x}}) &=\left( \begin{array}{c}
b_1 x_1^3 + b_2 x_1 x_2^2 + b_3 x_1\\ c_1 x_2^3 + c_2 x_2 x_1^2 + c_3 x_2
\end{array}\right) .
\end{aligned}
\end{equation}
Minimizing out the coefficients associated with the in-plane displacements, we find that 
\begin{equation}\label{eq:minKoit}
\min_{(b_i,c_i)} \mathcal{E}_{vK}( \tilde{\mathbf{u}}, v) =  
\frac{\mu}{12}lw \epsilon^2 \theta^2 \Big(q(\hat{\kappa}_1, \hat{\kappa}_2) + g_f \hat{\kappa}_1^2 \hat{\kappa}_2^2 \Big)
\end{equation}
where
$q(\hat{\kappa}_1,\hat{\kappa}_2) = 1 + 2 \hat{\kappa}_2 (\hat{\kappa}_2 +1)  + 2(\hat{\kappa}_1 + \hat{\kappa}_2)^2 + 2 \hat{\kappa}_1(\hat{\kappa}_1 - 1)$ and 
\begin{align}
g_f = \Big(\frac{l^4}{l^4 + (2/15) l^2w^2 +(7/240) w^4} \Big)\Big( \frac{1}{15} \frac{\epsilon^2 \tau^2}{(h/w)^4} \Big) 
. \label{eq:geomFact}
\end{align}
Here, $\hat{\kappa}_i = h/(2 \epsilon \tau) \kappa_i$ are the normalized principal curvatures.  Since $q(\hat{\kappa}_1,\hat{\kappa}_2)$ is minimized at $\hat{\kappa}_1 = 1/2 = -\hat{\kappa}_2$, the first term in \eqref{eq:minKoit} prefers the deformation of the wedge to be anti-clastic bending as shown in Figure \ref{fig:DesignRidge}(d).  In contrast, the second term in this energy prefers one of the curvatures to be zero as shown in Figure \ref{fig:DesignRidge}(e,f).  The balance is determined by the non-dimensional $g_f$ which incorporates all design variables.  It is easy to calculate that the exchange of stability happens at  $g_f = 384$.   Further, for $g_f \gg 384$, either $\hat{\kappa}_1 \approx 1/4, \hat \kappa_2 \approx 0$ (Figure \ref{fig:DesignRidge}(e)) or $\hat{\kappa}_1 \approx 0, \hat \kappa \approx -1/4$  (Figure \ref{fig:DesignRidge}(f)) .

\begin{figure}[!t]
\centering
\includegraphics[width =3in]{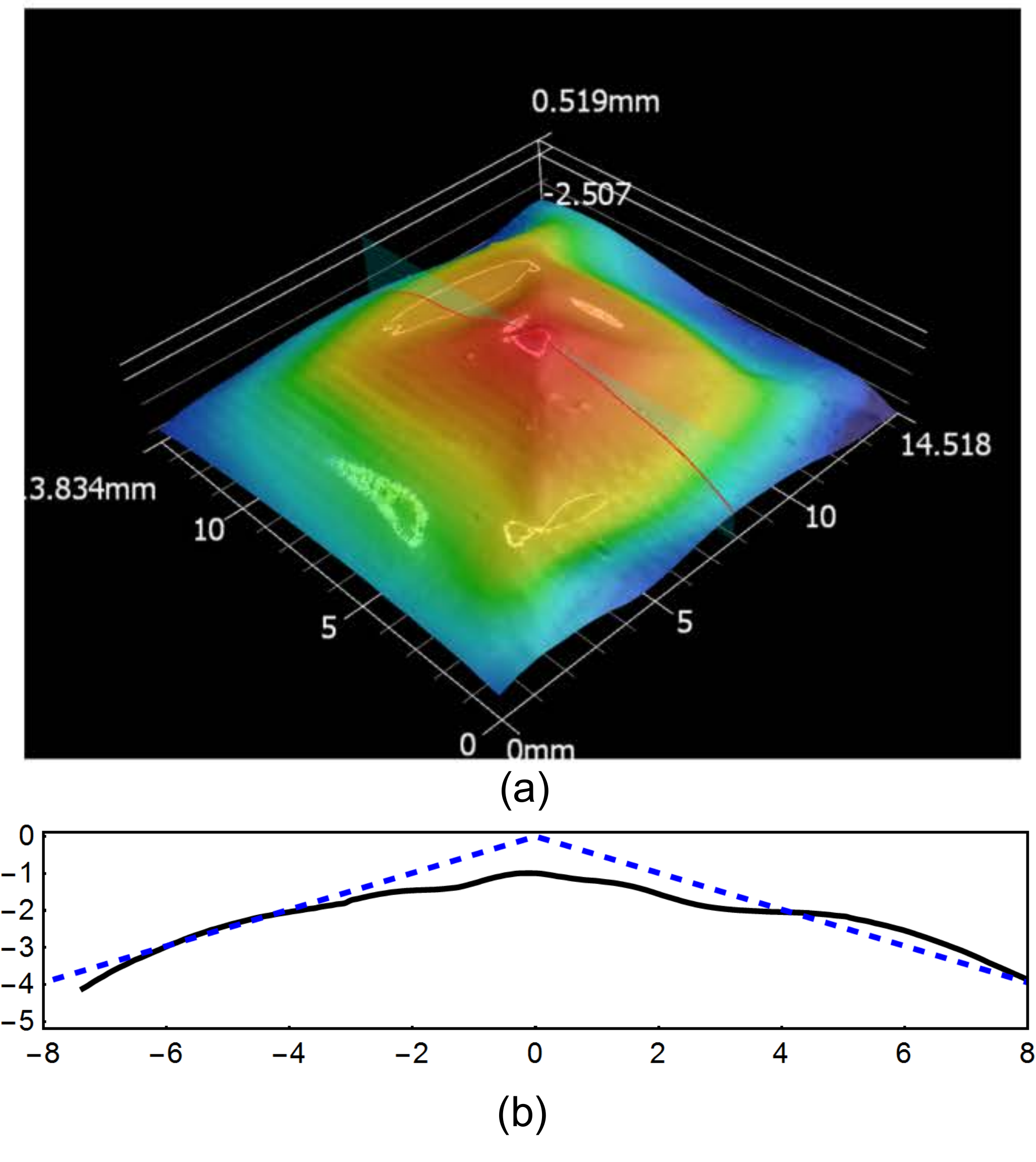}
\caption{Experimental observation of the bias pyramid.  (a) The actuated shape (the colors indicate elevation).  (b) A comparison of the observed cross-sectional slice (solid curve) with the predicted cross-section for the unbiased design (dashed blue line).}
\label{fig:BiasExper}
\end{figure}

\subsection*{Pyramid by design}  

We now apply this idea to the pyramid.   We imprint a monolithic sample with a director pattern shown schematically in Figure \ref{fig:DesignRidge}(g).  Note that this is the director pattern given by the nonisometric origami in Figure \ref{fig:DesignRidge}(a), except for the small wedge-shaped regions near the four interfaces where we introduce a small twist following the considerations in the previous section.      If $g_f \gg 384$, the wedges themselves favor being singly bent with either $\kappa_2 \approx 0 < \kappa_1$ (Figure \ref{fig:DesignRidge}(e)) or $\kappa_1 \approx 0 >  \kappa_2$ (Figure \ref{fig:DesignRidge}(f)).  Notice that the latter of these two configurations forms natural ridge consistent with the actuation of pyramid along an interface.  In contrast, the former imparts curvature along the interface and this is not consistent with the overall metric constraint.  Therefore, if $g_f \gg 384$, the  latter configuration ($\kappa_1 \approx 0 >  \kappa_2$) is preferred since these wedges are integrated into the overall pyramid design.  This breaks the up-down symmetry as shown in Figure \ref{fig:DesignRidge}(h).  Further, the ridges prefer curvature while the faces prefer to be flat.

We examine this experimentally in Figure \ref{fig:BiasExper}.  We consider a nematic elastomer with molar composition .75 RM82, 1 RM2AE and 1 EDT \cite{wetal_acsml_15} synthesized into a monolithic but patterned square film of dimensions $L = 17mm$ and thickness $h = 30 \mu m$.  We design the twist wedges for this pattern as in \ref{fig:DesignRidge}(g)  to bias the actuation up while choosing the width $w$ and twist angle $\tau$ so that negative Gauss curvature is suppressed.   In this direction, we note that the composition above gives\footnote{This value comes from the experimentally measured contraction data for this composition \cite{wetal_acsml_15}.} $r \approx 0.80$ and $\epsilon = 1-r^{1/2} \approx .11$ when the film is heated from room temperature to $160 \; C$.  Thus, we design the bias so that $\tau = \pi/12$ and width of the wedge is $w \approx 3.60 mm$.  Since $l \gg w$, this gives $g_f \approx 10000 \gg 384$.  

The measured shape on heating the sheet to $160 \; C$ is shown in Figure \ref{fig:BiasExper}.  The fabricated sheet actuates in the programmed direction when heated, and this is reproducible over dozens of heat cycles.  Further, as shown in the supplementary video, this bias is sufficiently robust that the pyramid promptly returns to its programmed direction even after being manipulated by hand into the "wrong" configuration.  Furthermore, the shape is in good agreement with the theoretical prediction.  The anisotropy parameter was measured to be $r = 0.8$ for this material and temperature change \cite{wetal_acsml_15}.  This leads to a prediction of the cross-section indicated by the dashed line in Figure \ref{fig:BiasExper}(b) for a pyramid without any bias and this agrees well with the observation.  The observed actuation is muted slightly in comparison since the twist nematic wedge regions induce a metric which differs by $O(\epsilon \tau)$ from the induced metric for the unbiased pyramid in these regions. 

\subsection*{Nonisometric origami as building blocks}

\begin{figure}[!t]
\centering
\includegraphics[width =3.5in]{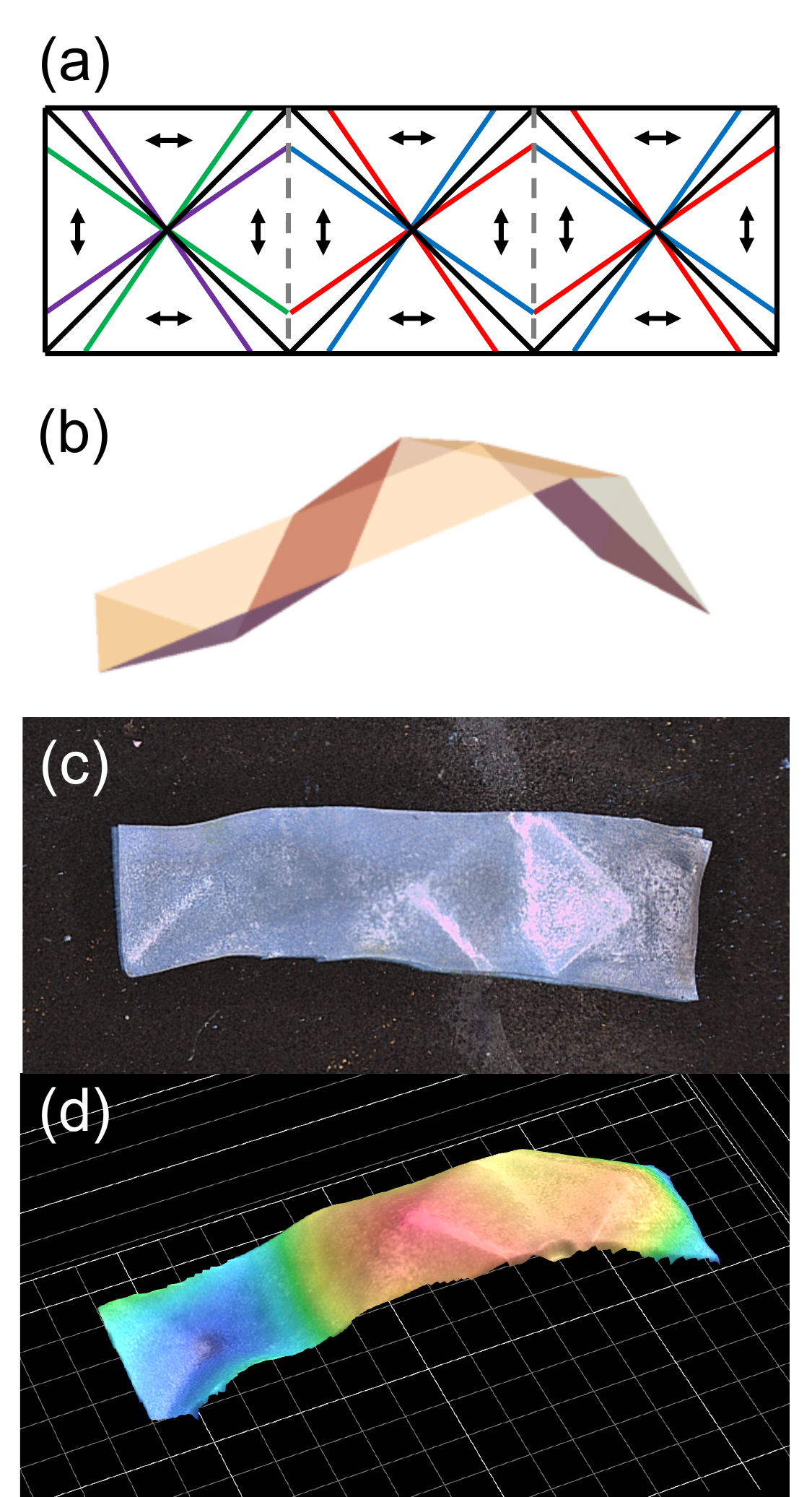}
\caption{Programmable actuation bias is demonstrated in a set of three joined pyramidal junctions.  (a) A schematic of the design.  For the center and right wedges, the twist is as in Figure \ref{fig:DesignRidge}(g).  For the left, the top and bottom director prescription is reversed (this is emphasized with the change in color).   (b) Intended shape, in which one pyramid actuates down and the remaining two actuate up.  The sheet is taken to rest on a flat surface.  (c) Top view of fabricated monolithic sample, patterned with three of the biased pyramidal building blocks from Figure \ref{fig:BiasExper}. Each building block is $5mm \times 5 mm$.  (d) Measured shape of sample upon heating shows that each pyramid actuates in its programmed direction. False color represents heights, with a span of 3 mm. }
\label{fig:BiasTriplet}
\end{figure}

We now show how to link simple nonisometric origami patterns together as building blocks to generate complex shape.  For the the efficacy of this design strategy, our ability to bias the simple building blocks to achieve a \textit{unique} actuation proves crucial.   We explain this by comparing experiments of designs composed of both biased and unbiased nonisometric origami building blocks.

We first consider the simplest possible design composed of building blocks of \textit{unbiased} nonisometric origami.  Figure \ref{fig:TwoPyramid} shows a design satisfying the metric constraint \eqref{eq:metric2D} with two pyramids linked across a region where the director is identical for each pattern. 
In particular, the sample in \ref{fig:TwoPyramid}(a) is patterned uniformly in the region where the building blocks are linked.\footnote{The grey dashed line in the figure is there simply to delineate the building blocks.}  Thus importantly, we expect the deformation to be uniform here (with no interface), and consequently, we expect the total deformation to be a complex three dimensional shape.  In other words, we expect that the pyramids do not deform in isolation but rather as an integrated structure.  These expectations are exactly realized in the experiments shown in Figure \ref{fig:TwoPyramid}(b).   However, since each pyramid can actuate either up or down, this design is multi-stable as observed.   We can rectify this using biased building blocks.

To highlight this, we now modify the design in \ref{fig:TwoPyramid}(a) by composing together three simple but \textit{biased} pyramid building blocks as depicted in Figure \ref{fig:BiasTriplet}(a).   For each building block, we prescribe twist nematic regions along the interfaces.  These are designed to break the up-down symmetry of the pyramid actuation, while ensuring that unwanted anti-clastic bending is suppressed.  Specifically, the center and right building block are designed to actuate up into a pyramid, while the left is designed to actuate down.  For the latter, we simply flip the top and bottom prescription of the director to achieve the opposite bias.  This design, in theory, results in a \textit{unique} energetically favorable actuation described by \ref{fig:BiasTriplet}(b) (this is simply a combination of the two actuations in \ref{fig:TwoPyramid}(c)).  The corresponding experiment in \ref{fig:BiasTriplet}(c-d) is in excellent agreement: The pyramids are biased, and the entire structure rotates over during the actuation to avoid bending the faces of uniform director. In doing all this, the patterned sheet achieves the predicted origami actuation. 

\section*{Conclusions}

\subsection*{Designing complex shapes} 

\begin{figure*}[!t]
\centering
\includegraphics[width = 6.5in]{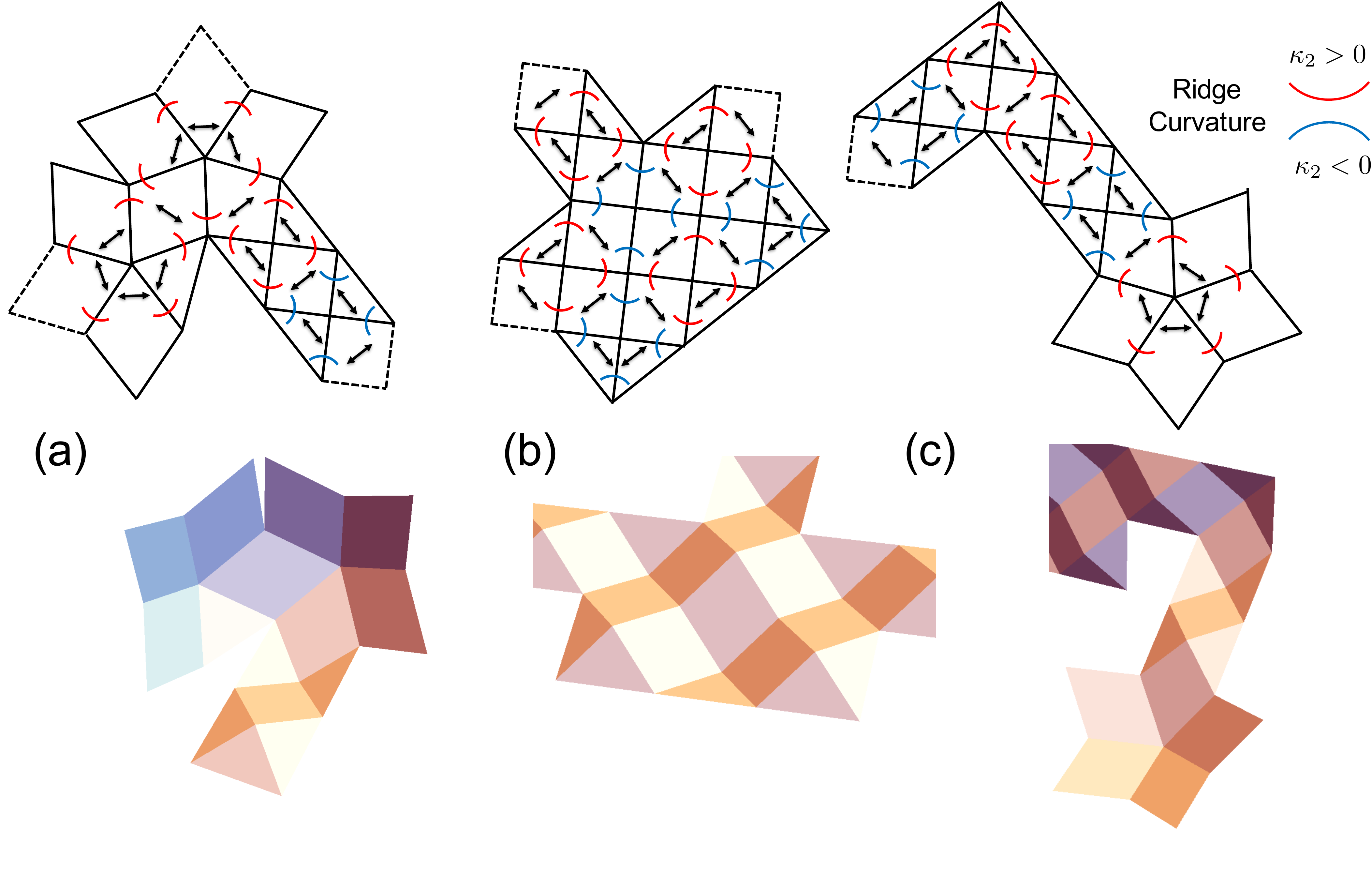}
\caption{Designing a humanoid soft robot by linking symmetric four-faced and five-faced building blocks. The figure described the design and actuation of (a) a portion of the head and neck region, (b) the body region, and (c) one of the arms. The colored arcs represent the designed curvature along the ridges of the origami; blue for a twist nematic wedge design to induce negative curvature transverse to the ridge, and red for a wedge design to induce positive curvature.}
\label{fig:ExplainSR}
\end{figure*}  
 
We have demonstrated experimentally two capabilities in actuating patterned nematic elastomer sheets: 1) Predictable three dimensional origami shape can be realized through the design of nonisometric origami building blocks.  2) These building blocks, when linked together, can be used to produce robust complex yet predictable, actuations.

This functionality enables a design landscape with striking possibilities.  We showcase this through a pattern to achieve, upon heating, the "humanoid" soft robot shown in Figure \ref{fig:SoftRobot}.   The design is composed of symmetric four and five-faced building blocks.  Each of the 27 total building blocks is independently compatible, and these are linked across regions in which the director profile is uniform.   For example, in the head and neck region depicted in Figure \ref{fig:ExplainSR}(a), we have two such linkages\textemdash one where a five-faced building blocks merges with a four-faced building block to form part of the neck and another where two five-faced building blocks merge to build a portion of the head.  To ensure that the design actuates the desired shape, we have designed a bias at each of the interfaces.  The blue arcs describe an appropriately designed wedge whose twist profile is prescribed at the top and bottom of the wedge as depicted in Figure \ref{fig:DesignRidge}(a).  This leads to symmetric pyramids which actuates up just as in \ref{fig:DesignRidge}(c).  In contrast, the red arcs describe a wedge design  \ref{fig:DesignRidge}(a) in which the top and bottom directors of the twist profile are reversed.   This leads to symmetric pyramids which actuate down. By controlling the shape and direction of actuation for these simple nonisometric origami building blocks and linking these together compatibly, we are able to systematically build a design to achieve the desired humanoid shape.

\section*{Acknowledgments}
This work draws from the doctoral thesis of P.P. at the California Institute of Technology.  P.P. is grateful for the support of the National Aeronautics and Space Agency through a NASA Space Technologies Research Fellowship.  K.B. and P.P. gratefully acknowledge the support of the Air Force Office of Scientific Research through the MURI grant no. FA9550-16-1-0566.  B.A.K. and T.J.W. gratefully acknowledge financial support of the Air Force Office of Scientific Research and the Materials and Manufacturing Directorate of the Air Force Research Laboratory.

\bibliographystyle{abbrv}
\bibliography{pbwbBib}

\newpage 
\section*{Supplementary videos}

The video nobias.mov shows the degeneracy of a pyramid designed with no bias (i.e., no twists near the ridges).  The pyramid can easily be manipulated to be stable in either the up or the down states.

The video bias.mov shows the robustness of the pyramid designed with a bias that is shown in Figure \ref{fig:BiasExper}.   The video shows that the pyramid promptly returns to its programmed direction even after being manipulated by hand into the "wrong" configuration.

\newpage 
\section*{Supplementary material on nonisometric origami}

\subsection*{Formulation of compatibility for general building blocks}

\begin{figure}
\centering
\includegraphics[width = 5in]{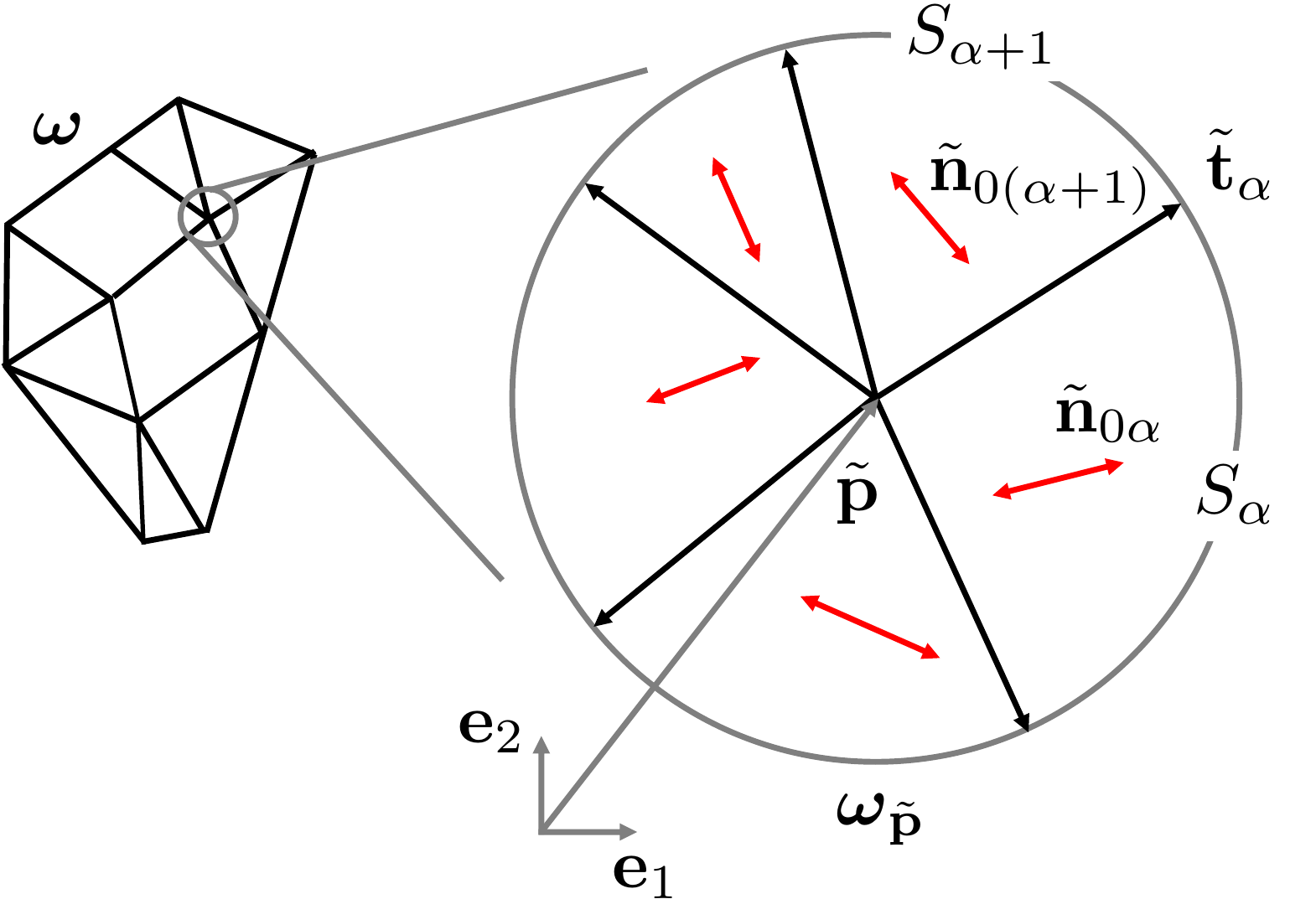}
\caption{At each vertex, the actuation induced shape change of the sectors surrounding the vertex must be comapatible for the pattern to be nonisometric origami.}
\label{fig:GenNonIso}
\end{figure}

Generic nonisometric origami designs are made up of building blocks where straight interfaces, which separate regions of distinct constant director, merge to a point as sketched in Figure \ref{fig:GenNonIso}.  (These interfaces, if designed appropriately, become the ridges of the origami structure upon actuation.)   Each of these building blocks must independently satisfy the metric constraint for the whole origami structure to be compatible upon actuation.  

To formulate this notion of compatibility of building blocks, we consider an initially flat $k$-faced building block $\omega_{\tilde{\mathbf{p}}} = \cup_{\alpha = 1,\ldots,k} S_{\alpha} \subset \mathbb{R}^2$ in which $k$ sectors $S_{\alpha}$ of piecewise-constant director merge to a point $\tilde{\mathbf{p}} \in \mathbb{R}^2$, which is in the interior of $\omega_{\tilde{\mathbf{p}}}$.  We let $\tilde{\mathbf{t}}_{\alpha} \in \mathbb{S}^1$ denote the outward tangent vector at the point $\tilde{\mathbf{p}}$ describing the interface between the sectors $S_{\alpha}$ and $S_{\alpha+1}$ of director $\mathbf{n}_{0\alpha}$ and $\mathbf{n}_{0({\alpha}+1)}$, respectively (where we define $\mathbf{n}_{0(k+1)} := \mathbf{n}_{01}$ and likewise for $S_{k+1}$).  The notation is also provided in Figure \ref{fig:GenNonIso}.

Given all this, there exists a design and deformation at the building block $\omega_{\tilde{\mathbf{p}}}$ which satisfies the metric constraint if and only if the collection of interfaces $\{ \tilde{\mathbf{t}}_{1}, \ldots, \tilde{\mathbf{t}}_{k} \}$ and directors $\{ \mathbf{n}_{01}, \ldots , \mathbf{n}_{0k} \}$ satisfies
\begin{equation}
\begin{aligned}\label{eq:nonIso}
&\mathbf{R}_{\alpha} (\bm{\ell}^{1/2}_{\mathbf{n}_{0\alpha}})_{3\times2} \tilde{\mathbf{t}}_{\alpha} = \mathbf{R}_{\alpha+1} (\bm{\ell}^{1/2}_{\mathbf{n}_{0(\alpha+1)}})_{3\times2} \tilde{\mathbf{t}}_{\alpha}, \\
&\qquad \alpha \in \{ 1,\ldots, k\}
\end{aligned}
\end{equation}
for some $\{ \mathbf{R}_{1}, \ldots, \mathbf{R}_{k}\} \in SO(3)$, where $(\bm{\ell}^{1/2}_{\mathbf{n}_{0\alpha}})_{3\times2} := r^{-1/6}(\mathbf{I}_{3\times2} + (r^{1/2} - 1) \mathbf{n}_{0\alpha} \otimes  \tilde{\mathbf{n}}_{0\alpha})$ for each $\alpha \in \{ 1,\ldots, k\}$. (Again, we define $\mathbf{R}_{k+1} := \mathbf{R}_1$.)   We have addressed the intimate connection between this compatibility and the metric constraint (1) elsewhere \cite{plb_pre_16, plb_arma_17}, as well as the justification of these configurations as designable actuation. Thus here, we focus on examples which highlight the richness of the design landscape for building blocks.

In this direction, we note that if the equations \eqref{eq:nonIso} are solved for some collection $\{ \mathbf{R}_{\alpha}\} \in SO(3)$, then any deformation $\mathbf{y} \colon \omega_{\tilde{\mathbf{p}}} \rightarrow \mathbb{R}^3$ of the form
\begin{equation}\label{eq:yDef}
\begin{aligned}
&\mathbf{y}(\tilde{\mathbf{x}}) = \mathbf{Q} \mathbf{R}_{\alpha} (\bm{\ell}_{\mathbf{n}_{0\alpha}}^{1/2})_{3\times2} (\tilde{\mathbf{x}} - \tilde{\mathbf{p}}) + \mathbf{y}(\tilde{\mathbf{p}}), \quad \tilde{\mathbf{x}} \in S_{\alpha}, \\
&\quad \alpha \in \{1,\ldots, k\}, \quad \mathbf{Q} \in SO(3) \quad \text{ and } \quad  \mathbf{y}(\tilde{\mathbf{p}}) \in \mathbb{R}^3
\end{aligned}
\end{equation}
realizes the metric on $\omega_{\tilde{\mathbf{p}}}$ and is, consequently, a candidate for the deformation of the building block upon actuation.  Furthermore, we can use the freedom afforded us in the rigid rotation $\mathbf{Q}$ and translation $\mathbf{y}(\tilde{\mathbf{p}})$ of the building block to attempt to link multiple such building blocks together to actuate complex shapes.  This framework permits a systematic investigation: (i) characterize the $\{\mathbf{t}_{\alpha}\}$, $\{\mathbf{n}_{0\alpha}\}$ and $\{\mathbf{R}_{\alpha}\}$ which satisfy \eqref{eq:nonIso}, (ii) use this characterization to build nonisometric origami building blocks described by deformations of the form in \eqref{eq:yDef}, and (iii) use these building blocks as building blocks to form more complex shapes. 

\subsection*{Some efforts on characterization}
For the characterization herein, we focus on the case that each of the directors $\mathbf{n}_{0\alpha} \in \mathbb{S}^2$ is planar  (i.e., $\mathbf{n}_{0\alpha} \cdot \mathbf{e}_3 = 0$ for each $\alpha$).  This encapsulates all such nonisometric origami building blocks which can be realized using current experimental techniques (e.g., the voxelation technique of Ware et al. \cite{wetal_science_15}).  We also assume that $\mathbf{n}_{0\alpha}$ and $\mathbf{n}_{0(\alpha+1)}$ are linearly independent (otherwise, the interface $\tilde{\mathbf{t}}_{\alpha}$ would be superfluous).  Finally, we assume that the sheet is under actuation (i.e., $r \neq 1$).

Under these assumptions, we can satisfy \eqref{eq:nonIso} only if the set of tangent vectors $\{\tilde{\mathbf{t}}_{\alpha} \}\in \mathbb{S}^1$ satisfies
\begin{equation}\label{eq:tangents}
\begin{aligned}
&\tilde{\mathbf{t}}_{\alpha} \in \text{span} \{ \tilde{\mathbf{n}}_{0\alpha} + \tilde{\mathbf{n}}_{0(\alpha+1)} ,  \tilde{\mathbf{n}}_{0\alpha} - \tilde{\mathbf{n}}_{0(\alpha+1)}\},\\
& \qquad \alpha \in \{ 1,\ldots ,k\}.
\end{aligned}
\end{equation}
In words, compatibility requires that the tangent vector at each interface bisect the two corresponding planar directors (up to a reflection of one or both of the directors).  This follows from taking the squared norm of both sides of \eqref{eq:nonIso}, and manipulating this quantity using the stated assumptions.

This condition on the interface tangents is necessary for a compatible nonisometric origami building block, but it is not sufficient.  For sufficiency, we need (additionally) to find a set of rotations $\{ \mathbf{R}_{\alpha}\} \in SO(3)$ which satisfies \eqref{eq:nonIso}.   In this direction, we note that all solutions to three-faced building blocks of nonisometric origami have been worked out explicitly in the thesis of Plucinsky \cite{p_thesis_2017}.   For instance, if we are given any three distinct planar director $\{\tilde{\mathbf{n}}_{01}, \tilde{\mathbf{n}}_{02}, \tilde{\mathbf{n}}_{03}\} \in \mathbb{S}^1$, then there are generically 16 nonisometric origami building block designs associated to these directors that are compatible for heating.  These design are bi-stable, (i.e., the pattern can actuate either up or design).  The formulas associated to the designs and actuations are cumbersome, but explicitly stated in Appendix A of the thesis and easily amenable to calculation (in mathematica, for instance).  We showcase the rich design landscape for three faced building blocks by considering one such example below.

Now in trying to generalize these results to ($k > 3)$-faced building blocks, there are important distinctions to be made.  Most notably, such building blocks, when compatible, are generically and non-trivially degenerate.  Specifically, if we can find a ($k>3$)-faced building block which has a solution to \eqref{eq:nonIso} upon actuation, then there are, in fact, continuous families of solutions.  The basic idea is that additional interfaces, beyond three, add degrees of freedom to the systems (i.e., we pick up an extra folding angle rotation for every additional interface).  These extra degrees of freedom allow for continuous families of solutions.  Indeed, recall that for the symmetric four-faced building block design, we showed that the interfaces can be continuously deformed from the (initially symmetric) pyramid to  generate an entire family of "metric invariant" actuations arising from this design (Figure 3(c) in the main text). We emphasize again that this is a generic fact of ($k > 3$)-faced building blocks (though we will refrain from supplying the formal argument here).   

From a design point of view, these result have important implications:  due to the degeneracies of the nonisometric building blocks, we need a strategy that breaks the up-down symmetry and distinguishes between various metric-invariant actuations associated to building blocks with more that three interfaces.  We believe that in tailoring the twist angles $\tau$ and the width of the wedges $w$, our strategy to bias the interfaces can enable this simultaneous functionality.  

\subsection*{Examples: General three-faced building blocks}
\begin{figure}
\centering
\includegraphics[width = 5.5in]{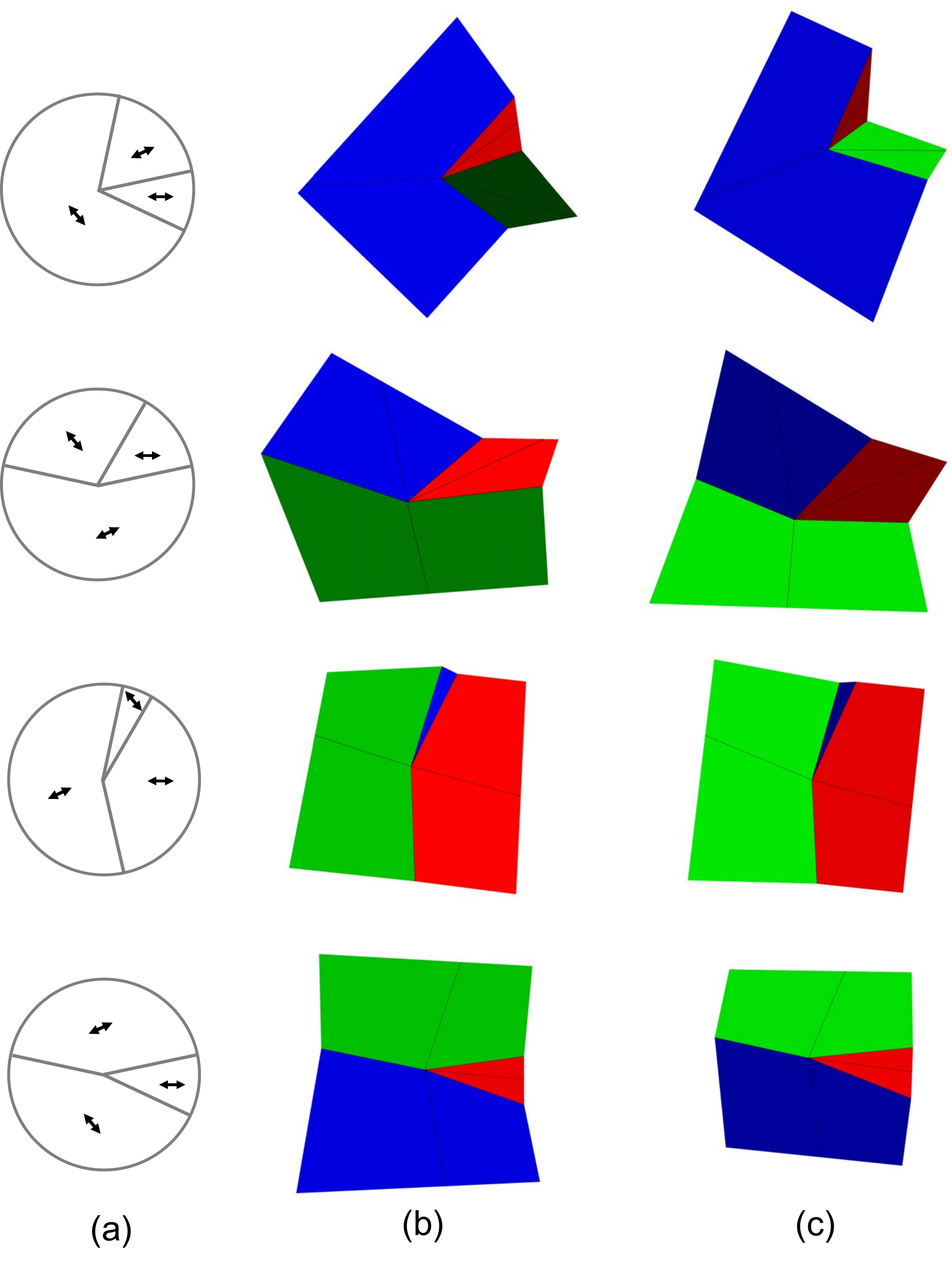}
\caption{(a) Director design for the building block.  (b),(c) The bi-stable actuations for $r = 0.5$.}
\label{fig:GenExample1}
\end{figure}

\begin{figure}
\centering
\includegraphics[width = 5.5in]{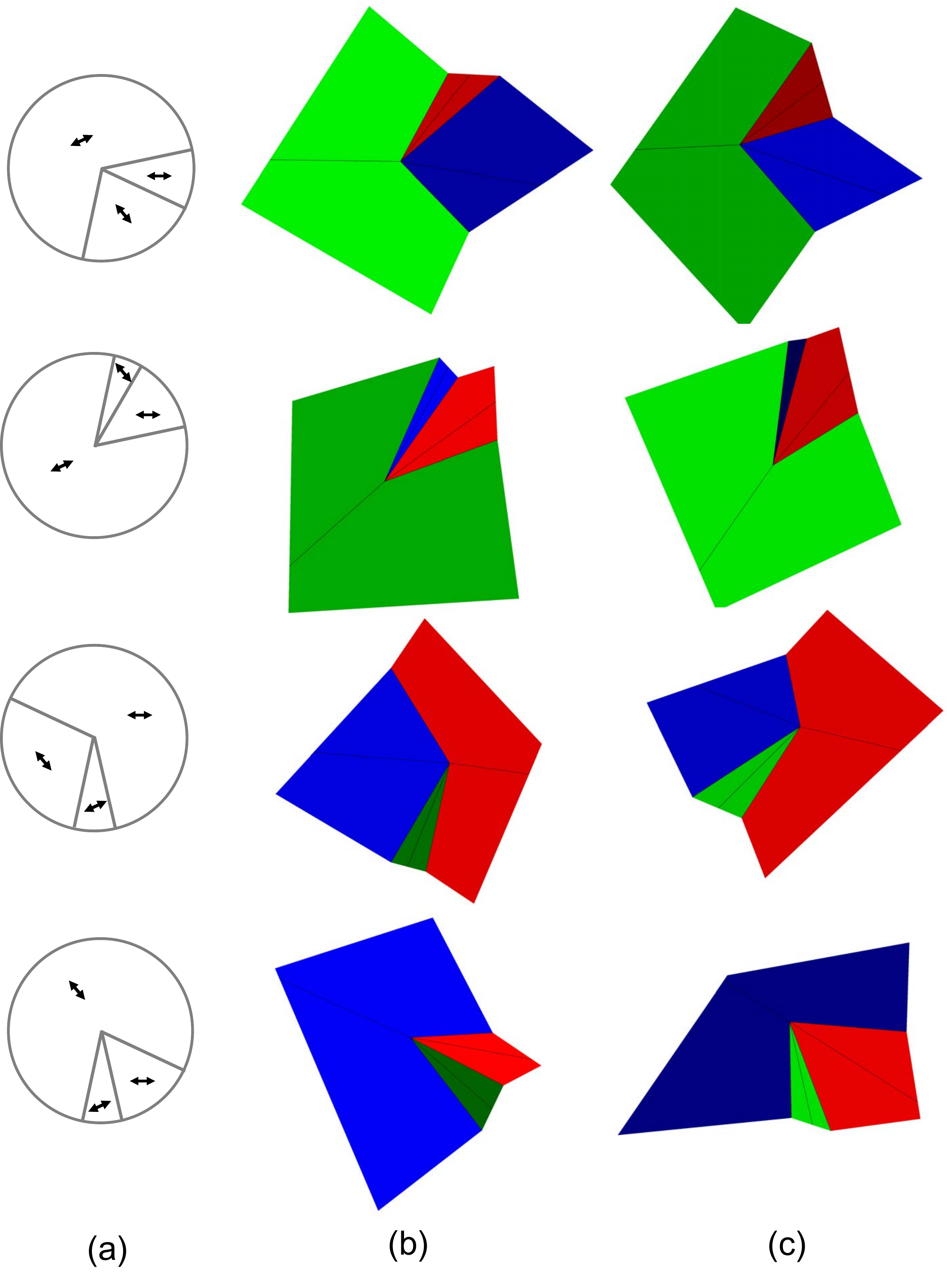}
\caption{(a) Director design for the building block.  (b),(c) The bi-stable actuations for $r = 0.5$.}
\label{fig:GenExample2}
\end{figure}

\begin{figure}
\centering
\includegraphics[width = 5.5in]{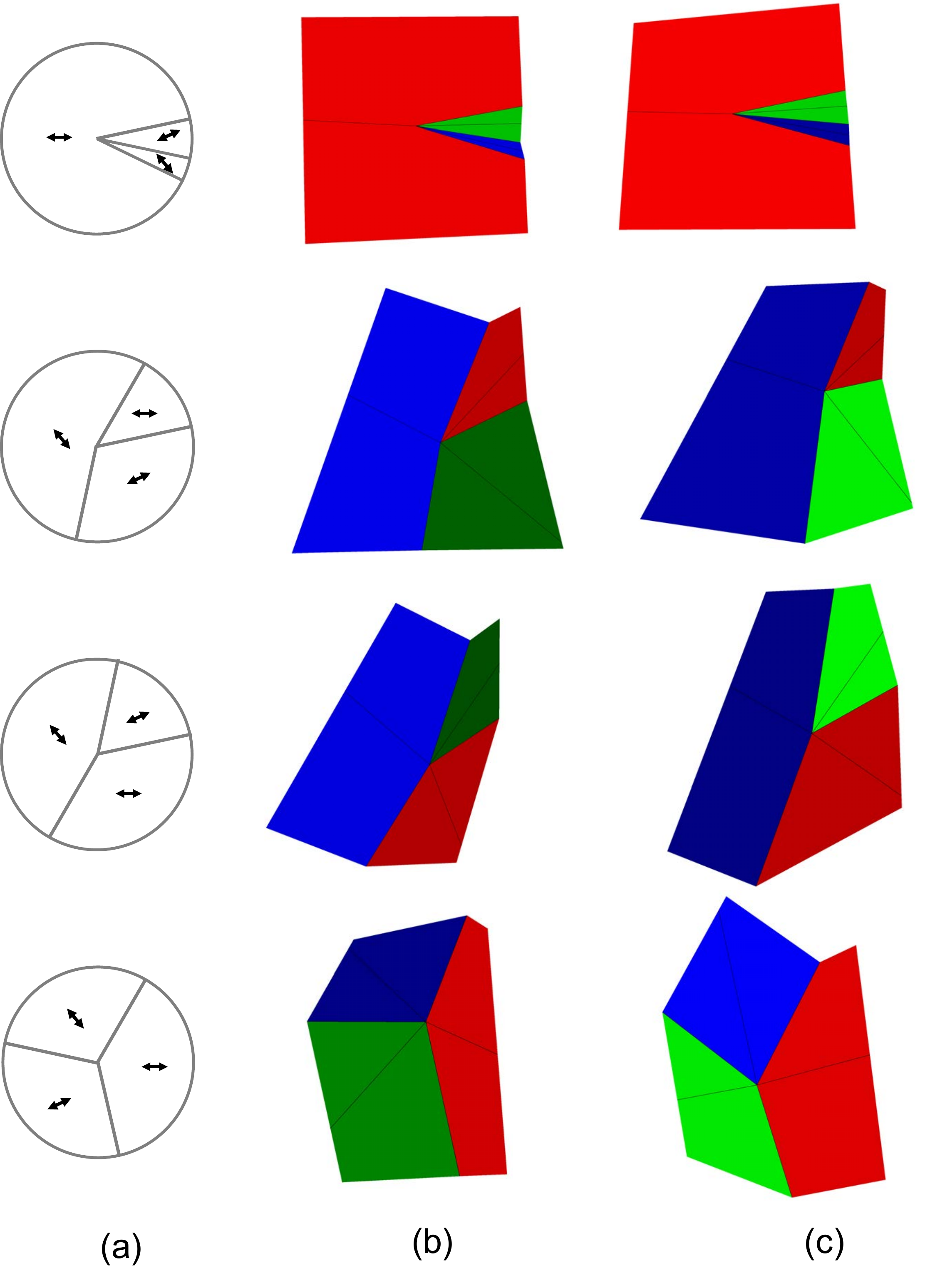}
\caption{(a) Director design for the building block.  (b),(c) The bi-stable actuations for $r = 0.5$.}
\label{fig:GenExample3}
\end{figure}

\begin{figure}
\centering
\includegraphics[width = 5.5in]{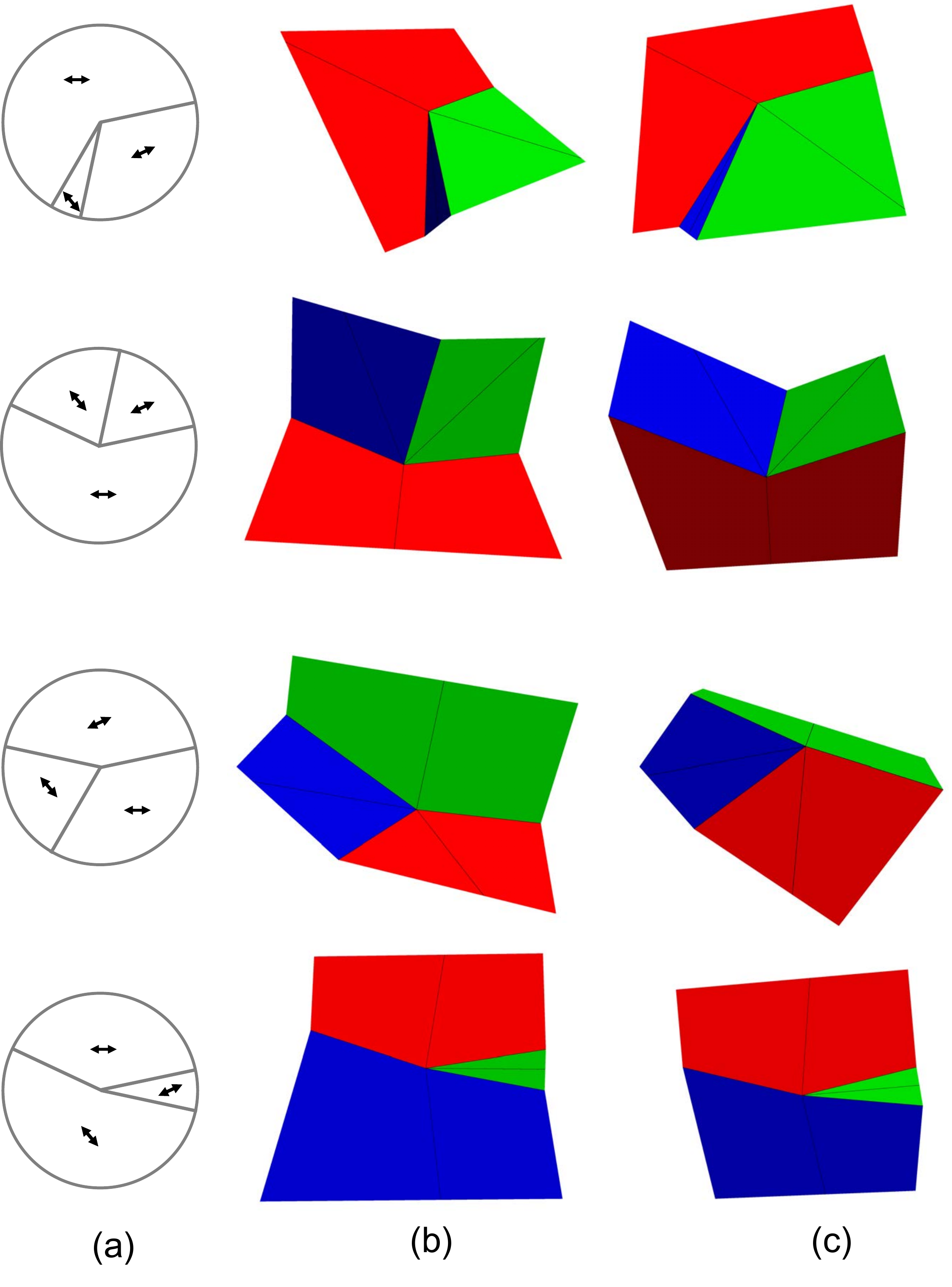}
\caption{(a) Director design for the building block.  (b),(c) The bi-stable actuations for $r = 0.5$.}
\label{fig:GenExample4}
\end{figure}

To highlight the richness of the design landscape for nonisometric origami building blocks, we consider a single set of directors, 
\begin{align}
\{ \mathbf{n}_{01}, \mathbf{n}_{02}, \mathbf{n}_{03}\} = \{ \mathbf{e}_1,\;\; \cos( \frac{5 \pi}{36}) \mathbf{e}_1 + \sin( \frac{5 \pi}{36}) \mathbf{e}_2, \;\; \cos( \frac{5 \pi}{18}) \mathbf{e}_1 - \sin( \frac{5 \pi}{18}) \mathbf{e}_2  \} ,
\end{align} 
and provide all the compatible nonisometric origami building blocks and corresponding actuations associated to this set.  There are 16 such compatible designs, and each design and actuation is provided in Figures \ref{fig:GenExample1}, \ref{fig:GenExample2}, \ref{fig:GenExample3} and  \ref{fig:GenExample4} (explicit formulas for the actuation are a direct result of Theorem A.3.6 in \cite{p_thesis_2017}).  The actuation parameter in all the examples shown is $r = 0.5$, but the solutions exist and are continuous for all $r \in (0,1]$.  The bi-stable solutions are reflections of each other.  In other words, the second solution is obtained by replacing the folding angles of the first with folding angles of the same magnitude but opposite sign.  The colors are associated to a particular director: red is $\mathbf{n}_{01} = \mathbf{e}_1$, green is $\mathbf{n}_{02} =  \cos( \frac{5 \pi}{36}) \mathbf{e}_1 + \sin( \frac{5 \pi}{36}) \mathbf{e}_2$, and blue is $\mathbf{n}_{03} = \cos( \frac{5 \pi}{18}) \mathbf{e}_1 - \sin( \frac{5 \pi}{18}) \mathbf{e}_2$.  

We emphasize again that all these examples correspond to a \textit{single set} of directors (notice the arrows in Figure \ref{fig:GenExample1}(a), \ref{fig:GenExample2}(a), \ref{fig:GenExample3}(a) and  \ref{fig:GenExample4}(a)).  If we change the set, then we would obtain a new collection of nonisometric origami building blocks, likely 16 but sometimes less if their is some symmetry in associated to the set of directors.  Thus, there are an infinite number of three-faced building blocks, and this is the simplest possible case.  There is still much to be explored in the direction of $(k > 3)$-faced building blocks.

\end{document}